\documentstyle[prl,aps,epsfig]{revtex}

\catcode`@=11
\def\citer{\@ifnextchar [{\@tempswatrue\@citexr}{\@tempswafalse\@citexr[]}}
 
\def\@citexr[#1]#2{\if@filesw\immediate\write\@auxout{\string\citation{#2}}\fi
  \def\@citea{}\@cite{\@for\@citeb:=#2\do
    {\@citea\def\@citea{--\penalty\@m}\@ifundefined
       {b@\@citeb}{{\bf ?}\@warning
       {Citation `\@citeb' on page \thepage \space undefined}}%
\hbox{\csname b@\@citeb\endcsname}}}{#1}}
\catcode`@=12

\def\reffi#1{\mbox{Fig.~\ref{#1}}}
\def\reffis#1{\mbox{Figs.~\ref{#1}}}

\def\citere#1{\mbox{Ref.~\cite{#1}}}
\def\citeres#1{\mbox{Refs.~\cite{#1}}}

\newcommand{\mste}{m_{\tilde{t}_1}}
\newcommand{\mstz}{m_{\tilde{t}_2}}

\newcommand{\msbe}{m_{\tilde{b}_1}}
\newcommand{\msbz}{m_{\tilde{b}_2}}

\newcommand{\MstL}{M_{\tilde{t}_L}}

\newcommand{\MsbL}{M_{\tilde{b}_L}}

\newcommand{\MsqL}{M_{\tilde{q}_L}}
\newcommand{\MsqR}{M_{\tilde{q}_R}}

\newcommand{\At}{A_t}
\newcommand{\Ab}{A_b}
\newcommand{\Xt}{X_t}
\newcommand{\Xb}{X_b}

\newcommand{\SU}{\mathrm {SUSY}}

\newcommand{\msbar}{$\overline{\rm{MS}}$}

\def\order#1{${\cal O}(#1)$}
\newcommand{\cp}{{\cal CP}}

\newcommand{\twol}{two-loop}

\newcommand{\fh}{{\em FeynHiggs}}

\newcommand{\subh}{{\em subhpole}}

\newcommand{\MW}{M_W}
\newcommand{\MZ}{M_Z}
\newcommand{\MA}{M_A}
\newcommand{\mh}{m_h}

\newcommand{\MH}{M_H}

\newcommand{\mt}{m_{t}}

\newcommand{\mq}{m_{q}}
\newcommand{\mb}{m_{b}}

\newcommand{\mgl}{m_{\tilde{g}}}
\newcommand{\sq}{\tilde{q}}

\newcommand{\sqe}{\tilde{q}_1}
\newcommand{\sqz}{\tilde{q}_2}

\newcommand{\Stop}{\tilde{t}}
\newcommand{\StopL}{\tilde{t}_L}
\newcommand{\StopR}{\tilde{t}_R}
\newcommand{\Stope}{\tilde{t}_1}

\newcommand{\Sbot}{\tilde{b}}
\newcommand{\SbotL}{\tilde{b}_L}
\newcommand{\SbotR}{\tilde{b}_R}

\newcommand{\Sbotz}{\tilde{b}_2}

\newcommand{\tsf}{\theta\kern-.20em_{\tilde{f}}}
\newcommand{\tsfp}{\theta\kern-.20em_{\tilde{f}\prime}}
\newcommand{\tsq}{\theta\kern-.15em_{\tilde{q}}}
\newcommand{\sw}{s_W}

\newcommand{\sweff}{\sin^2\theta_{\mathrm{eff}}}

\newcommand{\stt}{s_{\tilde{t}}}
\newcommand{\stb}{s_{\tilde{b}}}

\newcommand{\ctt}{c_{\tilde{t}}}
\newcommand{\ctb}{c_{\tilde{b}}}

\newcommand{\VL}{\left( \begin{array}{c}}
\newcommand{\VR}{\end{array} \right)}
\newcommand{\ML}{\left( \begin{array}{cc}}
\newcommand{\MLd}{\left( \begin{array}{ccc}}
\newcommand{\MLv}{\left( \begin{array}{cccc}}
\newcommand{\MR}{\end{array} \right)}

\newcommand{\tb}{\tan \beta}

\newcommand{\gev}{\,\, {\mathrm GeV}}

\newcommand{\BC}{\begin{center}}
\newcommand{\EC}{\end{center}}
\newcommand{\BE}{\begin{equation}}
\newcommand{\EE}{\end{equation}}
\newcommand{\BEA}{\begin{eqnarray}}
\newcommand{\BEAnn}{\begin{eqnarray*}}
\newcommand{\EEA}{\end{eqnarray}}
\newcommand{\EEAnn}{\end{eqnarray*}}
\newcommand{\non}{\nonumber}
\newcommand{\id}{{\rm 1\kern-.12em
\rule{0.3pt}{1.5ex}\raisebox{0.0ex}{\rule{0.1em}{0.3pt}}}}
\newcommand{\lsim}
{\;\raisebox{-.3em}{$\stackrel{\displaystyle <}{\sim}$}\;}
\newcommand{\gsim}
{\;\raisebox{-.3em}{$\stackrel{\displaystyle >}{\sim}$}\;}

\def\al{\alpha}

\def\als{\alpha_s}

\def\be{\beta}

\def\de{\delta}

\def\si{\sigma}

\def\Ga{\Gamma}

\def\De{\Delta}

\begin{document}

\twocolumn[\hsize\textwidth\columnwidth\hsize\csname
@twocolumnfalse\endcsname


\title{Do electroweak precision data and Higgs-mass constraints rule out\\
a scalar bottom quark with mass of ${\cal O}(5 \gev)$?}
\author{ M.~Carena$^{1}$, S. Heinemeyer$^2$, C.E.M.~Wagner$^{3}$
and G.~Weiglein$^4$ \\}
\address{
$^1$ Fermilab, Box 500, Batavia, IL 60510-0500, USA
\\
$^2$ HET, Physics Department, Brookhaven Natl.\ Lab., Upton, NY 11973, USA
\\
$^3$ High Energy Division, Argonne National Laboratory, Argonne,
IL 60439, USA\\
and the Enrico Fermi Institute, Univ.\ of Chicago, 5640 Ellis, Chicago,
IL 60637, USA\\
$^4$ Theoretical Physics Division, CERN, CH-1211 Geneva 23, Switzerland}
\maketitle
\begin{abstract}
We study the implications of a                               
scalar bottom quark, with a mass of \order{5\gev},           
within the MSSM. Light sbottoms may naturally appear         
for large $\tb$ and, depending on the                        
decay modes, may have escaped experimental detection.        
We show that a light sbottom cannot be                       
ruled out by electroweak precision data and the              
bound on the lightest $\cp$-even Higgs-boson mass.           
We infer that a light $\Sbot$ scenario requires a            
relatively light scalar top quark whose mass is typically    
about the top-quark mass.                                    
In this scenario the lightest                                
Higgs boson decays predominantly into $\Sbot$ pairs          
and obeys the mass bound $\mh \lsim 123 \gev$.
\end{abstract}


\vskip2.0pc]


 
New light particles, with masses of the order of the weak scale,
are an essential ingredient in any scenario beyond the standard
model (SM) that leads to an explanation of the large hierarchy 
between the Planck mass and the weak scale. 
Although no clear evidence of such a particle
has been reported so far, searches for new particles 
are usually
performed under model-dependent assumptions and hence the quoted
bounds may not be valid if these assumptions are relaxed. In 
particular, we shall investigate whether a light scalar bottom quark, 
$\Sbot$, with mass close to the bottom-quark mass, $\mb$,
is consistent with present experimental data~\cite{pdg}.
A light $\Sbot$ is most naturally obtained within
supersymmetric theories~\cite{mssm} for large values of 
$\tb$, as required in minimal SO(10) scenarios~\cite{SO10}.
Supersymmetric theories have received much attention in the last years since
they  provide an elegant way to break the 
electroweak symmetry and to stabilize the huge hierarchy between the GUT 
and the Fermi scales; they also allow for a consistent unification of the gauge
couplings. 
Supersymmetry (SUSY) predicts the existence of scalar
partners to each SM fermion, and spin-1/2 partners to the gauge and
Higgs bosons. 

Scalar particles, like the $\Sbot$, have been searched for at 
current and past collider experiments. Despite being in the mass 
reach of these colliders, they may have been overlooked for several 
reasons. Bottom squarks give only a tiny contribution to the inclusive
cross section for $e^+e^- \to$ hadrons, smaller than two percent of
the total quark contribution for five flavors of quarks,
and therefore
small compared to the experimental error in these measurements~\cite{pdg}.
Furthermore, due to a p-wave suppression of the fermion 
contribution to its decay width, a
$\tilde{b}\bar{\tilde{b}}$ resonance would be difficult to extract 
from background~\cite{Nappi}. 
Concerning the semileptonic decay of the $\Sbot$, 
$\Sbot \to c \; l$ + missing energy, if its branching ratio is small, 
for instance of about
the bottom quark one, the exclusion bound derived by the CLEO
collaboration does not apply~\cite{CLEO}. 
If, on the other hand, the light $\Sbot$ decays
into a light quark and missing energy, due to its small mass
and the small mass splitting between the $\Sbot$ and its decay
products, it cannot be detected through missing energy searches
in $e^+e^-$ or hadron colliders~\cite{pdg}.
If, instead, 
the $\Sbot$ decays fully hadronically 
with no missing
energy, 
it will remain undetected due to its small contribution to
the hadronic cross section at hadron and lepton colliders.
Finally, the presence of a light $\Sbot$ will slightly affect the extrapolated
value of the electromagnetic and strong
gauge couplings, $\alpha_{em}$ and $\alpha_s$,
at the scale $M_Z$: 
the variation induced on $\alpha_{em}(M_Z)$ 
is smaller than the
difference between the two most commonly used values 
of $\alpha_{em}(M_Z)$~\cite{jeg}. 
The variations 
of both $\al_{em}$ and $\alpha_s(\MZ)$
are smaller than the present error on 
the respective coupling~\cite{pdg}.


On the other hand, the hadronic observables measured with high precision
at the $Z$~peak at LEP1~\cite{lepewwg} impose tight and fairly 
model-independent
constraints on this kind of new physics, provided that the $\Sbot$
couples with sufficient strength to the $Z$. A necessary
condition for such a scenario within the minimal
supersymmetric standard model (MSSM) to be phenomenologically viable
is thus a relatively small coupling of the $\Sbot$ to the
$Z$~boson.
The  squark couplings to the 
$Z$ depend on the mixing angle, $\tsq$, 
\BEA
g_{Z\sqe\sqe} &\simeq& g \left( T_3 \cos^2\tsq
- Q_{\sq} \sin^2\theta_W \right) , \non \\
g_{Z\sqe\sqz} &\simeq& g \, T_3 \sin\tsq\cos\tsq , \non \\
g_{Z\sqz\sqz} &\simeq& g \left( T_3 \sin^2\tsq
- Q_{\sq} \sin^2\theta_W \right) ,
\EEA
where $\sin^2\theta_W \equiv \sw^2 = 1 - \MW^2/\MZ^2$; in the 
following the shorthand notation $s_{\tilde q} \equiv \sin\tsq$ and 
$c_{\tilde q} \equiv \cos\tsq$ is used.
In the particular case of the $\Sbot$, $Q_{\tilde{b}} = -1/3$, $T_3 = -1/2$,
and hence an exact cancellation of the coupling of the 
lightest $\Sbot$, $\Sbotz$, to the $Z$ is achieved
in lowest order when 
$\stb^2 = 2/3 \sw^2$, i.e.\ $|\stb| \approx 0.38$. 
Similarly, an exact cancellation for  
the lightest $\Stop$, $\Stope$, yields $\ctt^2 = 4/3 \sw^2$. 
For our conventions in the squark sector, see \citere{mhfd}.

Besides the constraints from the direct search and from
$Z$-peak observables for the $\Sbot$,
the considerable splitting between the
masses in the scalar bottom and top sector, which are necessary to avoid 
direct observation of at least one of these particles at LEP,
gives rise
to sensitive restrictions from virtual effects to electroweak precision
observables, e.g.~$\sweff$, $\MW$, $\Ga_l$, via contributions to the
$\rho$-parameter.  
Therefore, it is of interest to investigate whether a
$\Sbot$ almost mass-degenerate with the bottom quark  
is consistent with the strong
constraints from electroweak precision data. 
A further crucial question is whether a light $\Sbot$ scenario 
can give rise to a sufficiently large value for the lightest
$\cp$-even Higgs-boson mass in the MSSM in view of the bounds arising
from the Higgs searches at LEP.
The latter constraints have
meanwhile ruled out a considerable part of the parameter space, even in
the unconstrained MSSM (in which no assumptions about the underlying
SUSY-breaking mechanism are made)~\cite{lephiggs}. The
present bound on the SM Higgs mass from the direct search 
is $\MH > 113.3\gev$ at 95\% C.L.~\cite{lepchiggs}. The upper bound 
on the lightest $\cp$-even Higgs mass within the MSSM 
is $\mh \lsim 130\gev$ for $\mt = 175\gev$. This bound arises 
from the theoretical prediction of $\mh$ in the MSSM up to the two-loop
level~\cite{mhfd,mhrg}. 



As a first step in our analysis
we have calculated the production cross section for light scalar bottoms
as a function of the effective $Z\Sbotz\Sbotz$ coupling
(throughout this paper we use the tree-level notation for this
coupling, although it can be viewed as an effective coupling containing
loop corrections). As an additional scenario to the case where this
coupling precisely vanishes, we have taken the sbottom mixing in 
the range $|\stb| \approx 0.3$--0.45.
If the $\Sbot$ would decay with a small semileptonic decay width,
in a way similar to the bottom quark,
it would mainly affect observables associated with bottom
production, as discussed below. Analyzing the
corresponding effects on the relevant $Z$ peak observables, 
$R_b$, $R_c$, $R_l$, $A^b_{FB}$, $A_b$, $\Ga_{\rm had}$, $\Gamma_Z$
and $\si_{\rm had}$,
for $|\stb|$ = 0.3, 0.45
we find the
following results for the comparison of the data with the predictions, 
given in units of standard deviations:
$\de R_b = 0.40 \si (1.0 \si)$,
$\de R_c = -1.01 \si (-1.04 \si)$,
$\de R_l = 0.62 \si (1.08 \si)$,
$\de A_{FB}^b = -2.33 \si (-2.42 \si)$,
$\de A_b = -0.48 \si (-0.55 \si)$,
$\de\Ga_{\rm had} = 0.09 \si (0.57 \si)$,
$\de\Ga_Z = -0.85 \si (-0.43 \si)$,
$\de\si_{\rm had} = 1.87 \si (1.62 \si)$.
The values in brackets correspond to the SM predictions~\cite{lepewwg}.
The agreement of the predictions with the data improves
over the SM case for most observables.
Lowering $\als$ by $\approx 0.0018$ ($\approx 0.6 \si$)~\cite{pdg},
$R_l$, $\Ga_{\rm had}$, $\Ga_Z$ and $\si_{\rm had}$ 
would reach their SM values, 
whereas $R_b$, $R_c$, $A_{FB}$ and $A_b$ would to a good approximation 
keep the above improved values. Thus, a
small but non-vanishing coupling of the light $\Sbot$ to the $Z$ not only is
compatible with the hadronic observables at the $Z$ peak, but may even
slightly improve the agreement with the data.
Since the shifts discussed here are small, the overall quality of a
global fit to all data is expected to change only slightly. The same is
true if, alternatively, the sbottom decays only hadronically.

As a second step in our analysis,
we investigate the constraints from $\Delta\rho$
and the Higgs mass limit for the two cases:\\[-1.7em]
\begin{itemize}
\item[(I)]
Vanishing coupling of $\Sbotz$ and $\Stope$
to the $Z$~boson,
$\stb = \pm \sqrt{2/3}\, \sw$,
$\ctt = \pm \sqrt{4/3}\, \sw$.\\[-1.6em]
\item[(II)]
Small $Z\Sbotz\Sbotz$ couplings 
corresponding to the range of mixing
angles $|s_{\tilde b}| \approx 0.3$--0.45. 
No constraints on the $Z\Stope\Stope$
coupling are imposed.\\[-1.7em]
\end{itemize}
In the analysis below, $\msbz$ has been fixed to $4 \gev$, but 
varying this mass by a few GeV would not qualitatively change our results.
Since we also restrict $\stb$ as specified above, in principle
there are four more free
parameters left in the scalar bottom and top sector, $\msbe$, $\mste$,
$\mstz$ and $\stt$.
The relation between these parameters in the
mass-eigenstate basis and the ones in the basis of the current
eigenstates $\SbotL, \SbotR$, $\StopL, \StopR$ is given by the mixing
matrices
\BE
{\cal M}^2_{\sq}  =  
\ML
\MsqL^2 + \mq^2 + D_{\tilde q_L} & \mq X_q \\ \mq X_q &  
\MsqR^2 + \mq^2 + D_{\tilde q_R}
\MR
\EE
for $q = t, b$, and
$\Xt=\At-\mu \cot\beta$,  
$\Xb=\Ab-\mu \tb$. The
D-term contributions $D_{\tilde q_{L,R}}$
have not explicitly been written.
In the above,
$A_{t,b}$ denote the trilinear Higgs--$\Stop$, --$\Sbot$ couplings,
respectively, and $\mu$ is the Higgs mixing parameter.
SU(2) gauge invariance leads to the relation $\MstL = \MsbL$.
Thus only three of the four parameters $\msbe$, $\mste$, $\mstz$,
$\stt$ are independent.

Since the heavier $\Sbot$ has not been observed at LEP2,  and it
can in principle be produced in association with the lighter one, its
mass should be larger than (conservatively)
$\sim\,$200~GeV. Neglecting terms of order
$\msbz^2/\msbe^2$, the mass of the heavier $\Sbot$ is given
as
$\msbe^2 = \mb \Xb/(\stb\ctb)$.
In order to generate a sufficiently
large  value of $\msbe$, relatively large 
values of 
$X_b$ are required.
They can naturally be obtained
for values of $|\mu|$ and $A_b$ around the squark masses if
$\tb \approx |\stb\ctb| \msbe/\mb$, 
where $\mb \approx 3 \gev$ is the \msbar\ running
bottom mass at the weak scale. For heavy $\Sbot_1$ masses
of order 400 GeV and $\Sbot$ mixing angles of the cases (I, II),
this implies values of $\tb \gsim 30$.

Concerning the constraints from contributions of the $\Stop$--$\Sbot$
sector to $\Delta\rho$, the present data leave some room for a small
but non-zero 
contribution to $\Delta\rho$. We use $2 \times 10^{-3}$ as upper bound
for SUSY contributions~\cite{pdg}. We have checked that a limit on
$\De\rho^{\SU}$ as tight as $3 \times 10^{-4}$ does not qualitatively
change our results.

Regarding the Higgs mass constraints,
beyond the tree level, the main correction to $\mh$ stems from the 
$t$--$\Stop$ sector and, for large values of $\tb$, also from the 
$b$--$\Sbot$ sector.
For a light $\Stop$ and $\Sbot$ sector, the Higgs tends to be light.
For large values of $\tb$ and $\MA$, however, the Higgs may be heavy
enough to avoid LEP constraints, but tends naturally to be
in the range 110--120 GeV. 
Concerning the bounds obtained at LEP2, one should note that
the off-diagonal
term in the $\Sbot$ mass matrix of the order of the square
of the weak scale (i.e.\ a large value of $(\mu\tb)$)
results in a large coupling
of these sbottoms to the lightest $\cp$-even Higgs boson.
Therefore, for large $\tb$ and $\MA$ the width of its decay into 
sbottoms,
\BE
\Gamma(h \to \Sbotz\bar{\tilde{b}}_2) \sim 
G_F \sqrt{2} (m_b \mu \tan\be \stb \ctb)^2/
(8 \pi m_h) ,
\label{hsbsb}
\EE
will be much larger than the corresponding one into bottoms,
$\Gamma(h \to b \bar b) \sim G_F  \sqrt{2} (m_h m_b^2)/(4 \pi)$.

The limits from LEP will depend strongly on the decay modes of the
sbottoms. As a conservative bound, we adopt the present lower bound 
on the Higgs boson of the SM at 
LEP2, $\mh \gsim 113.3$~GeV~\cite{lepchiggs}. 
This is consistent with the assumption that
the light $\Sbot$ decay channels are similar to the bottom quark ones.
However, if it decayed fully hadronically with no missing energy or
into down (or strange) quarks and missing energy, 
considerably weaker Higgs mass bounds would be obtained.

For the case of a very light $\Sbot$, with a non-negligible component
on the left-handed $\Sbot$, the constraint from the $\rho$-parameter
demands a relatively light $\Stop$. 
The simultaneous requirement that the lightest $\cp$-even Higgs mass 
should be above the experimental bound 
leads to strong
restrictions in the $\Stop$ sector.
In the numerical analysis, 
we use the following parameters: $\mt = 174.3 \gev$, $\mb = 3 \gev$,
$\tb = 40$, $\MA = 800 \gev$, $\mgl = 200 \gev$, 
$\mu = \pm 250 \gev$, $M_2 = 200 \gev$. We have chosen a large value 
for $\MA$, yielding that
the upper bound for $\mh$ within this scenario is only weakly 
dependent on 
the actual value of this parameter~\cite{lepbenchmarks}. 
The dependence on $\mgl$, $\mu$ and $M_2$ is also weak.

The theoretical predictions for $\mh$ employed here are based on the
\twol\ results of \citeres{mhfd,mhrg,bse}, implemented in the 
programs \fh\ \cite{feynhiggs} and \subh\ \cite{mhrg,bse}. 
We have checked that the results for $\mh$ obtained with
the two programs are close to each other and
therefore  lead to similar  conclusions.
$\De\rho^{\SU}$, including leading \twol\
contributions~\cite{drho}, has been evaluated with \fh.

The analysis is performed for the cases (I, II) defined above.
It should be emphasized that, although case (I) seems highly
constrained, starting from the requirement of
a small $\Sbot_2$ mass and a vanishing coupling to the $Z$, 
and requiring the left-handed $\Stop$ mass to be larger
than the right-handed one, most
solutions to the precision observables and Higgs
mass constraints would lead to a small coupling of
the lightest $\Stop$ to the~$Z$. 

In \reffi{fig:mst1mst2} the allowed parameter regions for $\mste$ and
$\mstz$ for the cases (I)
and (II) are shown, obeying the $\mh$ and $\De\rho^{\SU}$ constraints.
For both cases a considerable part of the 
parameter space is consistent with the 
constraints. In case (I) the allowed regions are 
$70 \gev \lsim \mste \lsim 220 \gev$, 
$450 \gev \lsim \mstz \lsim 600 \gev$.
In case (II) the $\Stop$ masses obey the constraints for
$70 \gev \lsim \mste$ $\lsim 330 \gev$, 
$400 \gev \lsim \mstz$, and we considered 
values of $\mstz \le 1000 \gev$. 
%
\begin{figure}[ht]
\begin{center}
\mbox{
\psfig{figure=mst1mst2.eps,width=6cm,height=4.3cm}}
\end{center}
\caption[]{
Regions in the $\mste$--$\mstz$ plane for the cases (I) and 
(II), allowed by the requirements $\mh \gsim 113.3 \gev$ and 
$\De\rho^{\SU} < 0.002$.
(See text for the other parameters.)
}
\label{fig:mst1mst2}
\vspace{-1em}
  \end{figure}
%
%
\begin{figure}[ht]
\begin{center}
\mbox{
\psfig{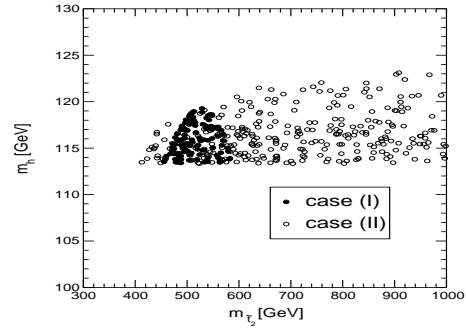}}
\end{center}
\caption[]{
Regions in the $\mstz$--$\mh$ plane for the cases (I) and 
(II), allowed by the requirements $\mh \gsim 113.3 \gev$ and 
$\De\rho^{\SU} < 0.002$.
(See text for the other parameters.)
}
\label{fig:mhmst2}
  \end{figure}
%
In  \reffi{fig:mhmst2} the allowed parameter regions for
$\mh$ are shown. In case (I) the lightest $\cp$-even Higgs will
always be lighter than 120 GeV, while in case (II) 
slightly larger values of $\mh$ can be obtained, $\mh \lsim 123$~GeV.
If the light sbottoms 
decay in a way similar to the $b$~quarks, this
offers good chances for the Higgs boson discovery at the Tevatron 
or the LHC, using 
its associated
production with the gauge bosons~\cite{TeVHiggs} or with the
top quark~\cite{Rainwater,Atlas}.

Scalar top masses below or about $100\gev$ are
constrained by LEP data. This mainly applies to case (II), in 
which no constraints on the $Z\Stope\Stope$ coupling were imposed, 
therefore allowing for larger contributions from 
$Z$ exchange to the $\Stope\Stope$ production cross section.
It follows from \reffis{fig:mst1mst2} and \ref{fig:mhmst2} that
only small changes would be obtained if this
bound were applied.

 Let us stress that the fine-tuning of the parameters
necessary to accomodate a $\Sbot$ mass of about $\mb$~is
about the same as the one necessary to realize the~SM 
with new physics arising at energies of about a few TeV. 
Concerning the mixing in the $\Sbot$ and $\Stop$ sector, the analysis of
case (II) shows that the mixing angles can be varied over a considerable
range, e.g.~$|\stb| \approx 0.3$--0.45, 
without leading to conflicts 
with the experimental constraints. The fact that not much fine tuning is
necessary is reflected in the large amount of experimentally
consistent models (see Figs.~\ref{fig:mst1mst2} and \ref{fig:mhmst2}).

In conclusion, a light $\Sbot$ within the MSSM cannot at present be
ruled out by the electroweak precision data and the
Higgs mass constraints from LEP2. Even in the most extreme case of
vanishing couplings of the lightest $\Stop$ and the lightest $\Sbot$
to the $Z$, an allowed parameter region within the MSSM is found,
resulting in an upper value for $\mh$, $\mh \lsim 120 \gev$, for 
$\mt = 174.3 \gev$.
If the light $\Sbot$ decays like a $b$~quark and has a small but
non-vanishing coupling to the $Z$~boson, this may even yield a slightly
better agreement of the $Z$ peak observables
with the experimental data than in the SM. 
In this case $\mh$ is restricted to be $\mh \lsim 123 \gev$.
An important finding in both cases is that 
the scenario with a $\Sbot$ 
almost mass-degenerate to the $b$~quark requires, in general,
also a light $\Stop$ whose mass is typically around the $t$~quark mass.
If it is light enough,
such a $\Stop$ should be accessible at Run~II of the Tevatron.
If the sbottoms decay similarly to $b$~quarks, these light
stops and sbottoms could contribute to the third-generation quark cross
sections, whereas the measured Tevatron 
cross sections are, in general, larger than the SM
expectations~\cite{hightopxsec}. 
Besides promising very interesting phenomenological implications for
Run~II of the Tevatron and for the LHC, a scenario with a light $\Sbot$
could also be studied in detail at the upcoming $b$~factories. 

\smallskip
G.W.\ thanks G.~Buchalla, A.~Dedes, H.~Dreiner, M.L.~Mangano
and D.~Zeppenfeld for interesting discussions. 
S.H.\ thanks H.~Eberl, S.~Kraml and C.~Schappacher for
technical support. C.W.\ and M.C.\
thank E.~Berger, B.~Dobrescu, B.~Harris, D.~Kaplan, H.~Logan,  
J.~Lykken, S.~Martin, K.~Matchev, S.~Mrenna, U.~Nierste, M.~Schmitt,
Z.~Sullivan, T.~Tait, D.~Wackeroth
and G.~Wolf for stimulating discussions and comments.
Work supported in part by 
US DOE, Div.\ of HEP, Contr.\ W-31-109-ENG-38.






\begin{references}

\bibitem{pdg} Part. Data Group,
              {\em Eur.\ Phys.\ Jour.\ }{\bf C15} (2000) 1.

\bibitem{mssm} H.E.~Haber and G.~Kane, 
               {\em Phys. Rep.} {\bf 117} (1985) 75;
               H.P.~Nilles, 
               {\em Phys. Rep.} {\bf 110} (1984)~1.



\bibitem{SO10}  
M.~Olechowski and  S.~Pokorski, {\em Phys. Lett.} {\bf B214} (1988) 393;
B.~Anantharayan, G.~Lazarides and Q.~Shafi, {\em Phys. Rev.} {\bf D44} 
(1991) 1613.



\bibitem{Nappi} C.R. Nappi, {\em Phys. Rev.} {\bf D25} (1982) 84.

\bibitem{CLEO} CLEO Collab., V. Savinov et al., hep-ph/0010047.



\bibitem{jeg} S.~Eidelman, F.~Jegerlehner, {\em Z.\ Phys.}
{\bf C67} (1995) 585;
M.~Davier, A.~H\"ocker, {\em Phys.\ Lett.}
{\bf B435} (1998) 427.

%

\bibitem{lepewwg} 
The LEP collaborations, LEP EWWG and SLD,
{\tt lepewwg.web.cern.ch/LEPEWWG/Welcome.html}.

\bibitem{mhfd} S.~Heinemeyer, W.~Hollik and G.~Weiglein,
                    {\em Phys. Rev.} {\bf D58} (1998) 091701;
                    {\em Phys. Lett.} {\bf B440} (1998) 296;
                     {\em Eur. Phys. Jour.} {\bf C9} (1999) 343.


\bibitem{lephiggs} P. Bock et al., 
                   CERN-EP/2000-055.

\bibitem{lepchiggs} 
A.~Read, LEPC presentation, 20.07.2000,\\
{\tt lephiggs.web.cern.ch/LEPHIGGS}.



\bibitem{mhrg}  M.~Carena, M.~Quir\'os and C.E.M.~Wagner,
                      {\em Nucl. Phys.} {\bf B461} (1996) 407.


\bibitem{lepbenchmarks} M.~Carena, S.~Heinemeyer, C.E.M.~Wagner and
                       G.~Weiglein, 
                       hep-ph/9912223.
%

\bibitem{bse} 
M.~Carena, H.E.~Haber, S.~Heinemeyer, W.~Hollik, C.E.M.~Wagner
and G.~Weiglein, {\em Nucl. Phys.} {\bf B580} (2000) 29.

\bibitem{feynhiggs} S.~Heinemeyer, W.~Hollik and G.~Weiglein,
                    {\em Comput. Phys. Commun.} {\bf 124} (2000) 76.

\bibitem{drho}     A.~Djouadi, P.~Gambino, S.~Heinemeyer, W.~Hollik,
                    C.~J\"unger and G.~Weiglein,
                    {\em Phys. Rev. Lett.} {\bf 78} (1997) 3626;
                    {\em Phys. Rev.} {\bf D57} (1998) 4179.



\bibitem{TeVHiggs} M.~Carena, S.~Mrenna and C.E.M.~Wagner,
{\em Phys. Rev.} {\bf D60} (1999) 075010;
{\em Phys. Rev.} {\bf D62} (2000) 055008;
M.~Carena, H.E.~Haber, J.~Conway, J.~Hobbs (conveners),
Higgs Working Group of the Tevatron Run~II 
Workshop, 
hep-ph/0010338.

\bibitem{Rainwater} J.~Goldstein, C.S.~Hill, J.~Incandela,
S.~Parke and D.~Rainwater, hep-ph/0006311.

\bibitem{Atlas} ATLAS Collab.,
report CERN/LHCC/99-15 (1999).

\bibitem{hightopxsec} 
F. Abe et al., CDF Collab., {\em Phys. Rev. Lett.} {\bf 80} (1998) 2779 
and {\bf 71} (1993) 2400;  
B. Abbott et al., D0 Collab.,                     
{\em Phys. Rev. Lett.} {\bf 83} (1999) 1908 
and {\bf 84} (2000) 5478. 

\end{references}
\end{document}